# Optical Micro- and Nanofiber Pulling Rig

J. M. Ward[1], A. Maimaiti[1,2], Vu H. Le[1] and S. Nic Chormaic[1]

[1]Light-Matter Interactions Unit, OIST Graduate University, 1919-1 Tancha, Onna-son, Okinawa 904-0495, Japan
[2]Physics Department, University College Cork, Cork, Ireland

We review the method of producing adiabatic optical micro- and nanofibers using a hydrogen/oxygen flame brushing technique. The flame is scanned along the fiber, which is being simultaneously stretched by two translation stages. The tapered fiber fabrication is reproducible and yields highly adiabatic tapers with either exponential or linear profiles. Details regarding the setup of the flame brushing rig and the various parameters used are presented. Information available from the literature is compiled and further details that are necessary to have a functioning pulling rig are included. This should enable the reader to fabricate various taper profiles, while achieving adiabatic transmission of ~ 99% for fundamental mode propagation. Using this rig, transmissions ranging from 85-95% for higher order modes in an optical nanofiber have been obtained.

## I. INTRODUCTION

Evanescent waves in optical waveguides have received great attention recently due to the advantage of optical coupling with systems such as microresonators,[1-3] cold atoms,[4-8] their use in optical sorting,[9,10] their ability to form microcavities themselves,[11-13] and to support cavities.[14-16] Of the many available waveguide types, the tapered optical fiber - otherwise known as the optical micro- or nanofiber (MNF) - is an important device in these research fields because it is relatively easy to fabricate, produces a strong evanescent field, and can be directly integrated into existing fiber optic systems. The main general requirements for an MNF are (i) a uniform waist with an optical wavelength scale diameter and (ii) taper transitions that satisfy the adiabatic condition,[17,18] i.e., the fibers should have low loss. The most common way of fabricating such a tapered fiber is by heating and stretching a section of standard optical fiber in a flame.[19-36] The flame can either be stationary,[24,28,33] brushed,[18-21,23,25,29,31,34-36] or one can use a simulated flame brush,[26-27,32] which relies on a stationary flame but oscillating stages. Most flame setups use either an oxy-butane[33,34] or oxy-hydrogen torch. Low loss fibers down to 50 nm have been made using the self-modulation flame method.[22,38] Alternative heat sources include a focused $CO_2$ laser, which is scanned along the fiber[39-43] or a stationary $CO_2$ laser focused onto the fiber via a diffractive optical element.[44] Microfurnaces in the form of electric strip heaters,[45-49] and sapphire tubes[50,51] have also been used. In the case of sapphire tubes, the tube is heated using a $CO_2$ laser.

Each of these heat sources has its own distinct advantages and disadvantages; for example, with the $CO_2$ laser system a stable and well-controlled heat source is available. However, it is difficult to directly heat the fiber with the $CO_2$ beam when the diameter of the fiber is less than 1 μm due to the inverse square relationship between fiber radius and heating for a $CO_2$ source. In contrast, flame heat sources follow an inverse relationship between heating and fiber radius. As a result, the minimum achievable taper diameter when using a CO2 laser is larger than that for a flame.[39-44] On the other hand, electric strip heaters and microfurnaces can produce submicron-sized fibers, but difficulty in engineering the heater's hotzone for the fabrication of microstructured fibers is a critical disadvantage. Though the basic flame pulling rig system is one of the most well-known methods for achieving arbitrary biconical MNF shapes, it is considered a challenging task because of the technical difficulties in maintaining uniformity and adiabaticity of the taper due to flame instabilities caused by gas regulation and environmental factors, such as air flow, during the tapering process. However, by proper control of the flame shape and the flame-brush movement one can obtain tapers with a high degree of accuracy with regards to the profile and complexity,[31,35] while, at the same time, maintaining low transmission loss.

In order to maintain a high optical transmission along an MNF, the transformation of the fiber mode along the taper region must be adiabatic[19]. However, there are applications where the nanofiber length must be short and adiabatic over a broad wavelength range.[52] In some cases, even nonadiabatic or asymmetric tapers are needed in order to ensure that mode beating occurs.[53-56] Recently, efficient propagation of higher order modes through an MNF[33,36,57] has become of interest, especially in the field of fiber-based atom manipulation and trapping.[58-61] The fabrication of the aforementioned tapered fibers therefore requires a flexible pulling rig system. There are many papers in the literature (and some research theses) which describe experiments that use tapered fibers made using the flame brushing technique; however, the published works give few details about the actual experimental setup of the pulling rig and the specific parameters used. Here, we aim to summarize the key points of the flame brushing method by combining the information already available in the literature with the missing, but essential, details in order to enable the reader to build a basic rig and to fabricate a variety of tapered fiber shapes with very high adiabaticity. We provide a schematic explanation of the setup, detail all the components and parameters used, and discuss the common pitfalls. Exponential and linear taper profiles were fabricated, and their adiabaticity and uniformity were tested. Afterwards more complex fiber profiles were made by combining two different taper profiles. Finally, the higher order fiber mode transmission for an adiabatic tapered MNF is demonstrated. The transmission typically lies in the 85-95% range for higher order modes coupled into an MNF fabricated from an 80 µm diameter few-mode fiber. This transmission is at least 2.5 times as high as that previously reported in the literature for the same fiber.[33]

## II. EXPERIMENTAL SYSTEM

### A. System design

There are three different ways in which flame brushing rigs are usually set up for MNF fabrication. These rely on (i) a stationary flame with stacked pulling stages,[26,27] (ii) a stationary flame with independent pulling stages,[32] or (iii) independent pulling stages and a moving flame.[19-21,23,25,29,31,34-36] The three types of flame brushing rigs are shown schematically in Figure 1.

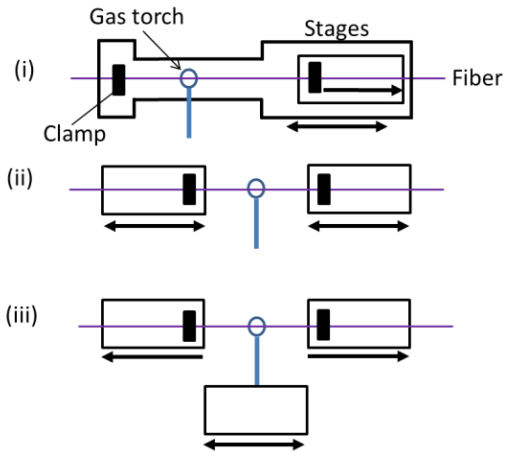

FIG. 1. (i) A stationary flame with stacked pulling stages, (ii) a stationary flame with independent pulling stages, and (iii) independent pulling stages and a moving flame. Double headed arrows indicate oscillating motion.

In the first setup, the flame brushing rig is composed of two linear translation stages, one stacked on top of the other. The flame is kept stationary under the fiber while the bottom stage performs a bidirectional motion which simulates a moving flame. The top stage, on the other hand, moves in only one direction and pulls on the fiber. This method is original; however, there may be some difficulties implementing such a setup. For example, there are greater loads on the bottom stage, so that larger forces are required to achieve a fast turning motion. These increased forces may cause the inertia of the top stage to affect the motion of the bottom stage. Pitch, yaw and roll of the stages can also cause bending of the fiber and induce transmission losses.[26,27] This setup can be used in another way whereby the two fiber ends are moved in the same direction such that the fiber is being pushed into the flame from one side while being pulled out on the other side with a greater speed.[31,37] In Ref. 37 an asymmetric pulling rig is described. One end of the fiber is fixed and the other end is pulled on by a motor. Similarly, a symmetric pull can be simulated by adding a constant velocity term to the flame motion;[31] this transformation moves the coordinates from the lab frame of reference to the fiber's reference frame.

The second possible design uses two independent stages and a stationary flame. The two stages are controlled to synchronously perform bidirectional motion while the flame is kept stationary beneath the fiber. During the fabrication process, flame brushing is done through the reciprocal movements of these two stages, while the fiber is elongated by increasing the separation between them. This setup removes mechanical coupling between the stages, but requires that the stages be individually aligned. Also, since both stages now oscillate, there is a higher degree of control needed to ensure minimum jerking or bending of the fiber at the turning points. Consequently, in one example of such a rig, a special iterative self-learning algorithm was implemented[32] to minimize the deviation from the set movements. These requirements may cause some difficulties for developing a control program.

The final option is to use three translation stages. One stage is used to oscillate the flame position along the fiber, while two stages are used to pull on either end of the fiber. The independence of each stage and the unidirectional motion of the pulling stages make this setup easier to control while alignment issues are no more severe than for other methods.[32] This setup also relaxes the requirements on the pulling stages, so that simple and relatively cheap stepper motors can be used. The disadvantage is that the flame is moving and there may be some deformation of its shape. If other air currents are removed this deformation should be small and consistent. To overcome all obstacles mentioned above, three linear motorized translation stages were used in the rig setup that we have implemented.

### B. Stage details and setup

A schematic of the experimental setup used in our design is shown in Figure 2. Two motorized translation stages (A) are used for pulling/stretching the optical fiber. They are Newport XMS100 series linear motor stages, with 100 mm travel distance. These stages have high accuracy which is achieved by an encoder with 80 nm repeatability.

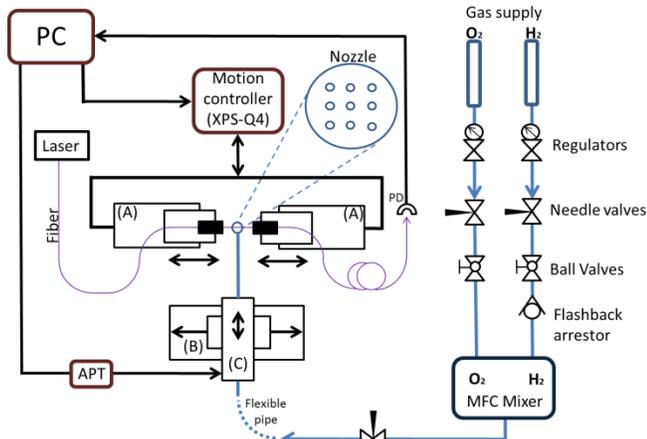

FIG.2. A schematic of the pulling rig system. MFC:Mass-flow controller, APT: stepper motor controller, (A) XMS linear motor stage, (B) ILS-LM linear motor stage, (C) stepper motor stage, PD: photodiode.

The gas burner nozzle is mounted onto a manual 1D micrometer stage (not shown in the schematic) which allows for the vertical position of the flame to be adjusted. The manual stage is, in turn, fixed on top of two motorized translation stages. The top stage (C) is a 1D stepper motor (Thorlabs APT) and is used to move the flame to a suitable position under the optical fiber and to remove the flame after pulling is complete. The bottom stage (B) is another linear motor translation stage (Newport ILS-LM series) which provides the linear motion of the nozzle along the fiber. This stage has high acceleration and can handle greater loads than the XMS series. All the linear motor stages are controlled by an XPS-Q4 controller which receives commands via a PC running a LabVIEW program.

### C. Fiber holding system

A pair of fiber clamps (Fushikara) was fixed onto each of the XMS linear motor translation stages. These clamps are used in fusion splicers and can usually be purchased separately. They provide a strong clamping force without bending the fiber and are available for fibers with cladding diameters of 125 μm and 80 μm (outer jacket diameters of 250 μm and 180 μm). The 180 μm fiber clamps can also be used to hold 250 μm fibers. Alignment of the fiber clamps is critical for low loss transmission and can be done with the aid of a straight edge, a spirit level and a keen eye. Care should be taken when tightening bolts as this will cause the position of the mounts to move.

Any dirt or external impurity on the fiber surface results in an abrupt drop in transmission and can cause the taper to have non-uniformities in its profile. After the acrylic jacket is stripped from the fiber, the fiber must be cleaned with acetone to remove inorganic matter and any remaining acrylic powder. A lens cleaning tissue or cotton buds soaked in acetone and repeatedly wiped along the stripped section of fiber removes most of the dirt. After the fiber is mounted into the clamps a digital microscope with a magnification of 250 is used to examine the fiber before pulling. Any remain particles can be removed using a clean acetone wipe. Even with rigorous cleaning some small particles may remain out of view; however, the cleaning procedure ensures transmission between 95% and 99%.

A clean working environment, such as a covered area with a fan and a filter, is required. The fan should be turned off during pulling to avoid air flow, but otherwise it should remain on to create a region of positive pressure in the enclosure containing the rig set-up.

Another issue to consider is the fact that, when the fiber is supplied, it is usually tightly wound around a drum and when it is unspooled it retains some of this curvature. When the fiber is clamped into the mounts it is straightened under tension. After heat is applied the fiber can relax and may bend slightly; therefore, de-stressing may be required before use.[30] Nonetheless, we found that, even if this slight bending occurs, overall transmissions in the 95% to 99% range are still possible.

### D. Gas System

The gas burner for this setup uses a premix of clean hydrogen and oxygen to avoid pollution of the tapered fiber. An oxy-butane mixture can also be used[33,34] although higher temperatures can be achieved with an oxy-hydrogen flame with lower gas flow rates, thus putting less gas pressure on the fiber. If the gas mixture of the oxy-butane torch is not correct then carbon soot can form on the fiber.[62] Surface contamination could lower the light transmission and the thin waist can melt in vacuum if too much light is absorbed. In order to achieve a repeatable and stable gas flame, mass flow controllers (MFCs) for hydrogen and oxygen are used (model FCST1000 (M) L, Fujikin). The controllers regulate the gas flow with an accuracy of 1% throughout the entire range of flow. Two gas regulators are connected to the gas source to provide accurate gas flow to the MFC. After the MFCs the gases are filtered and mixed before being sent to the gas burner via ¼ inch steel tubing. In order to achieve a full stoichiometric combustion of the gases and to eliminate any unused gas flow from the nozzle, a hydrogen:oxygen gas flow of 2:1 is used, i.e., in our setup a ratio of $H_2$ = 74 sccm to $O_2$ = 37 sccm (standard cubic centimeters per minute) of mixed gas was conveyed to the nozzle. The $O_2$ content can be increased fractionally (~ 1%) to give an $O_2$ rich gas mix which ensures a clean burn and removes additional organic compounds. However, the atmospheric $O_2$ around the flame is probably sufficient for this. The gas flow rate influences the flame temperature; a high flow rate produces a hotter flame which can cause the formation of β-cristobalite when heated above 1500°C.[63,64] If the glass is cooled quickly (e.g. by moving the flame away) then these crystals can be preserved and this reduces the optical quality of the glass.[63,64] The glass fiber, however, must be heated above the annealing temperature so that its viscosity is lowered sufficiently to allow deformation, but not heated above the softening temperature so that the glass does not sag under its own weight. This temperature range is between approximately 1200°C and 1500°C.[64] In effect, the total flow rate should be kept below 130 sccm to avoid losses.[32] After the MCF, the mixed gases are sent to the gas burner through ¼ inch steel tubing.

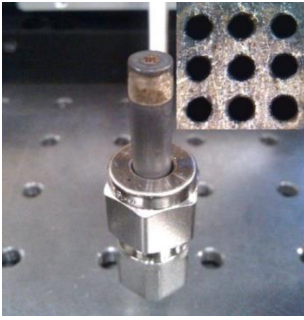

FIG. 3. Picture of the gas burner. The burner is made from a ¼ inch copper tube. A cap is brazed onto one end of the tube and an array of holes is drilled into the cap. The inset shows the array of holes, each hole is 200 μm in diameter.

The gas burner is made from a ¼ inch copper tube, one end of which is sealed by brazing a cap onto it (see Figure 3). A series of holes is drilled into the end cap - each hole has a diameter of 200 μm and there are nine holes in a 3 x 3 configuration with a total area of 1 mm². Other hole geometries may be preferable, such as a rectangular shaped array with the long axis perpendicular to the fiber. The small hole size prevents flame flashback to the fuel supply tubing. The optimum hole size is determined by the so called "quenching distance", which is the minimum size of hole that permits the flame to propagate. From the literature,[62] the quenching distance of hydrogen is 0.5-0.7 mm, therefore the required maximum hole size should be smaller than 0.5 mm to avoid any flashback-originated discontinuity of the gas flow. The copper nozzle was fitted to a ¼ inch steel tube using a right angle Swagelok connector and was clamped to the translation stages and connected to the output of the MFC via the flexible hose. A piezoelectric ignition system was used to generate a spark to ignite the gas. The nozzle was positioned so that the fiber was aligned above the center of the hole array. The distance from the top of the nozzle to the fiber is important; if it is too close then the fiber will be too hot and will glow white. We found the optimal distance by positioning the flame far below the fiber and gradually decreasing the distance until the fiber becomes hot enough to allow elongation without the fiber breaking or slipping through the clamps. The distance between the nozzle and the fiber was set to 2.2 mm in all of the experiments. This distance depends on the gas flow rates used and the shape of the hole array on the nozzle. In our setup, the gap could be changed by ±250 μm before we observed significant loss due to under or overheating. To be clear at this point, we define the hotzone as the region marked out as the flame transverses the fiber. The actual heated region on the fiber is much smaller and is the heated area on the fiber directly above the flame at any time. The size of the heated region will depend on the speed of the flame, the flow rate of the gases, the position of the fiber in the flame and the diameter of the fiber. Ref.32 gives a detailed discussion on how these parameters affect the final shape of the taper and uniformity of the taper waist. In Section IV we give details of appropriate flame speed

### E. LabVIEW control and run sequence

A LabVIEW program was written and used to control the translation stages. This program sends velocity, acceleration and position commands to the XPS-Q4 controller which drives the three linear translation stages. The program also commands the stepper motor, which positions the flame under the fiber. Before starting the pull, a single mode laser at a wavelength of 780 nm was coupled to one end of the fiber while the output end was connected to a photodiode (Thorlabs DET10A). The voltage signal from the photodiode was recorded on a digital oscilloscope (Agilent DSO5054A in high resolution mode). When the LabVIEW program activates the system, the stepper motor moves the flame to a position directly under the clamped fiber. There is a two second delay before the ILS-LM linear motor stage oscillates the flame at the prescribed velocity along the fiber over the desired length, which defines the hotzone. At the same time, the XMS linear motor translation stages pull the two ends of the fiber in opposite directions. The number of trips that the flame makes and the size of the hotzone depend on parameters such as the required taper profile, pull speed, and pull length. These details are discussed in the following sections. When the pull is finished the stepper motor removes the flame from underneath the fiber.

### III. ADIABATIC CONDITION

A tapered fiber can only be produced with high transmission if it fulfills the minimum adiabatic condition.[17,18] This requires that the tapering angle should be small enough to prevent the fundamental mode coupling to other, undesired, modes or radiation modes. The maximum taper angle which still allows adiabatic transmission is defined by:

$$\Omega_z = \frac{r(z)}{z_b} = \frac{r(z)(\beta_1 - \beta_2)}{2\pi}. \tag{1}$$

Here, $\Omega_z$ is the local half taper angle, $r(z)$ is the local radius of the taper transitions, $\beta_1$ and $\beta_2$ are the local propagation constants of the propagating mode and the nearest symmetric mode, and $z_b = 2\pi/(\beta_1 - \beta_2)$ is the beating length. For a low loss fiber, the local taper length scale $z_t \gg z_b$, where $z_t = r(z)/\Omega_z$ .

For a single-mode fiber (SMF) which supports only the fundamental, $HE_{11}$, mode the taper angle can be relatively large due to the propagation constant difference between two neighboring modes. In the case where the fundamental mode is excited at the input fiber, taper angles greater than the maximum angle in Equation 1 cause loss due to coupling to higher order modes. For fibers which support more than one mode, i.e. higher order modes which are generated at the input rather than those which arise due to an excessively steep nonadiabatic taper angle, the required angle for an adiabatic taper is smaller due to the decreased separation of the propagation constants between the modes.

The adiabatic condition for the fundamental mode has been studied extensively over the years.[17,18,65,66] Recently, attention has been given to the adiabatic transmission of higher order modes.[33,36,57] Petcu-Colan et al.[57] and, later, Frawley et al.[33] showed that higher order modes can efficiently propagate in linear taper profile MNFs if they are fabricated from optical fibers with reduced cladding diameters. The work by Ravets et al.[35] includes a detailed discussion about the evolution of the fiber modes during tapering. The paper quantifies the intermodal energy transfer of the fundamental mode in 125 μm diameter fiber and the practical limits of the tapering angle are determined.[35] Adiabaticity of higher order modes as a function of taper angle for 80 μm and 50 μm diameter fiber is discussed elsewhere. [33, 36]

## IV. EXPERIMENTAL RESULTS FOR A TRAVELING FLAME

### A. Exponential taper

When the flame is scanned back and forth along the fiber it creates a hotzone region of length $L$. Normally, $L$ is a function of the fiber elongation, $d$. When $L$ is a constant, such that $L = L_0$, the fiber radius as function of $d$ is expressed as:[20,21]

$$r_w(d) = r_0 e^{\left(-\frac{d}{2L}\right)}, \qquad (2)$$

where $r_0$ is the radius of the untapered fiber. $L_0$ is termed the constant hotzone. The local radius of the taper transitions can be used to define the fiber profile and it is given by

$$r(z) = r_0 e^{\left(-\frac{z}{L}\right)}. \qquad (3)$$

Here, $z$ is a coordinate along the fiber so that the length of the taper transition is $z_0 = d_0/2$, where $d_0$ is the total pulling length (i.e. the maximum elongation of the fiber). There is no time variable in the above equations; however, the glass drawn from the hotzone should be done at a rate that is significantly slower than the speed of the flame. The velocity of the flame, $V$, is as fast as 4 mm/s. Note that speeds much greater than this do not give enough time for the flame to heat the fiber unless the flame is hotter. Typically, the flame speed was set to 3 mm/s and the velocity of each pulling stage was set to 0.2 mm/s. The acceleration of the pulling stages was set to 1 mm/s$^2$ - this value was selected in order to ensure that there was a gentle initial pull on the fiber. The acceleration of the flame translation stage was set to 800 mm/s$^2$ in order to ensure that the flame spent as little time as possible at the turning points.

There is some play in these numbers, but these values worked well. For gas flow rates and flame position used, we found that by keeping the ratio of the pulling stage speed and the flame speed at 15 we were able to increase the flame speed up to 10 mm/s. At higher speeds the taper would break.

The number of oscillations (trips), $N$, that the flame stage makes depends on the pull length, the hotzone size, the speed of the flame and the pull rate, $P_r$. For example, for $d_0 = 40$ mm (20 mm/stage), $L = 4$ mm, $V = 4$ mm/s and $P_r = 0.4$ mm/s (i.e. 0.2mm/s per stage) the number of trips, $N = 50$.

As a demonstration, an exponential taper MNF was made from SMF (core and cladding diameter of 4.4 μm and 125 μm, respectively, for Thorlabs 780HP fiber) and its profile was compared against Equation 3. Laser light at a wavelength of 780 nm was coupled into the SMF; at this wavelength the fiber supports only the $HE_{11}$ mode. The transmitted power and the taper profile are shown in Figure 4(a). An overall transmission of ~ 99% for $HE_{11}$, was obtained. The fiber profile was measured under a microscope and showed good agreement between the measurement size and that predicted by Equations 2 and 3.

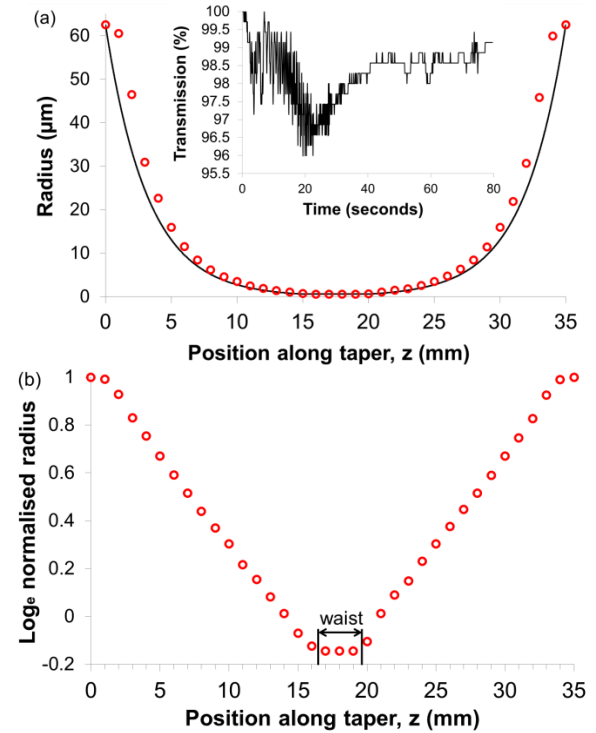

FIG. 4(a) Exponential tapered profile. The corresponding measured waist diameter was ~ 1.1 μm. The pull length was 30 mm and the flame travel distance was set to 3 mm. The solid line is a fit using Equation 3 with $L$ = 3.2 mm. The limited resolution (1 μm) of the microscope and finite width of the flame accounts for the difference between the set flame travel distance and $L$. Inset: Light transmission through the MNF. (b) The normalized measured radius plotted on a log scale.

The measured profile in Figure 4(a) is normalized and plotted on a log scale in Figure 4(b). The length of the hotzone can be estimated from the length of the waist region and from the slopes of the transitions using Equation 3. The length of the hotzone was measured as 3.1 mm and 3.2 mm from the slope and the waist, respectively.

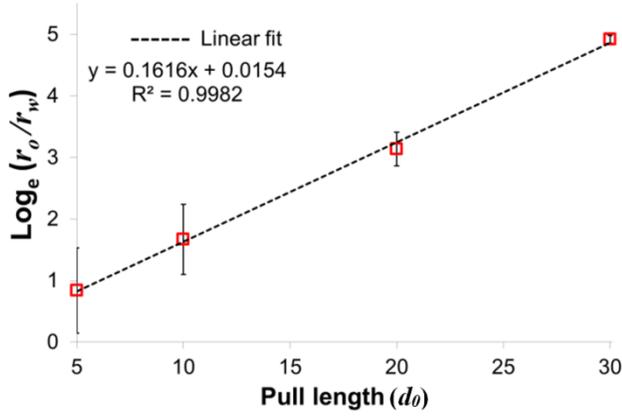

FIG. 5. The average waist radius for four different pull lengths, plotted as $ln(r_0/r_w)$. The slope of the linear fit is an estimate of the hotzone $L$, such that the slope is $L/2$. Ten MNFs were made for each pull length. The error bars show the standard deviation of the measured fiber waist.

Alternatively, the length of the hotzone on the fiber can be estimated by plotting $ln(r_0/r_w)$ against $d_0$,[21] where the slope = $L/2$, see Equation 2. The fiber waist radius, $r_w$, was measured for four different pull lengths. Ten fibers were made for each pull length and the average radius is plotted against $d$ in Figure 5. The resulting slope gives a hotzone of $L = 3.1$ mm , which is in good agreement with the scan length of the flame thus verifying that the traveling flame indeed produces the prescribed hotzone. The error bars in Figure 5 show the standard deviation of the measured fiber waist. Only the waist diameter was measured and the average shape was not determined. There were no fiber breakages during the 50 pulls shown in Figure 5; however, this does not mean that breakages never happens. Occasional breakage is usually due to damage to the fiber during stripping, a change in the gas flow rate, a change in the flame shape/position, or excessive dirt on the fiber.

## B. Linear taper

In experimental studies using MNFs, fibers with linear taper transitions have some substantial advantages. For example, the varying hotzone method has more flexibility to control the taper angle, uniformity, and length of the taper waist. One can also design more complex taper shapes by adding up different linear profiles or mixing the linear and exponential profiles in one taper. The idea of producing an arbitrarily shaped taper fiber and the associated computational model were first published by Birks and Li.[21] More complex shapes, including a taper that has a sinusoidal variation along the waist, have been demonstrated recently.[31] For creating arbitrary profiles, the so called "reverse problem" [21] is useful. In this method, the shape of the taper profile is designed in advance and is used as the input information. The computation framework calculates the hotzone size, $L(d)$, as a function of elongation. For a complete taper profile, some parameters should be provided such as the radius of the untapered fiber $r_0$, the length of the uniform taper waist, $l_w$, the final radius of the uniform taper waist, $r_w(d_0)$, and $r(z)$. Calculation steps are as follows: first, the length of the taper transition, $z_0$, is determined by solving the equation:

$$r(z - z_0) = r_w . \qquad (4)$$

Second, the solution of the volume law[21] yields the formula for the hotzone length, $L(z)$, as point $z$ is pulled out of the hotzone:

$$L(z) = \frac{(r_0)^2}{(r(z))^2} L_0 - \frac{2}{(r(z))^2} \int_0^z (r(z))^2 \, dz. \qquad (5)$$

The initial value of the variable hotzone, $L_0$, was extracted from Equation 5 such that $L(z_0) = l_w$ at $z = z_0$ and

$$l_w = \frac{(r_0)^2}{(r(z_o))^2} L_0 - \frac{2}{(r(z_o))^2} \int_0^{z_o} (r(z))^2 \, dz . \qquad (6)$$

Once $L(z)$ is known, Equation 5 with $L_0$ can be determined from Equation 6, and the elongation as a function of profile, $d(z)$, can be extracted from the inverted form of the distance law:[21]

$$d(z) = 2z + L(z) - L_0. \qquad (7)$$

Finally, the hotzone size as a function of elongation, $L(d)$, is calculated from the distance law by substituting the $z(d)$ function which has just been found:

$$L(d) = d + L_0 - 2z(d). \qquad (8)$$

The total pulling length is determined by the equation:

$$d_0 = 2z_0 + l_w - L_0. \qquad (9)$$

In Figure 6, the hotzone size as a function of elongation, $L(d)$, is plotted for two different taper profiles. An overall trend of reduced hotzone size for increased fiber elongation is evident.

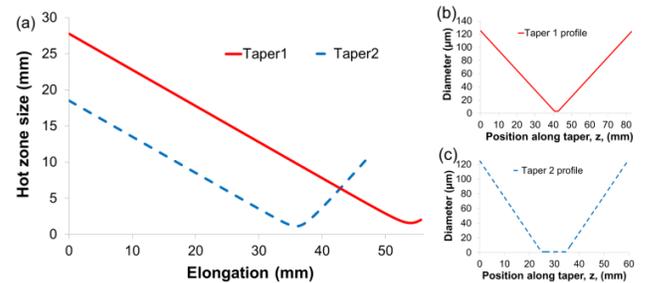

FIG. 6(a) The hotzone size as a function of elongation $L(d)$ for two different tapers. (b) Taper 1: $l_w = 2$ mm , $r_w = 3\ \mu m$ and $\Omega_z = 1.5$ mrad. (c) Taper 2: $l_w = 10$ mm, $r_w = 1\ \mu m$ and $\Omega_z = 2.5$ mrad.

With the moving flame setup it is possible to convert $L(d)$ into a position and velocity profile for the flame. This information can be tabulated and used to command the translation stages. To make the conversion one must know the pull rate of the translation stages, $P_r$, the number of trips, $N$, the time per elongation, $T$, and the pull interval, $I_p = d_0/N$. $P_r$ is simply the rate at which the stages pull apart. $I_p$ is the amount by which the fiber is elongated for each trip of the flame. The time per elongation $T = I_p/P_r$, is the time between each turning point. For a given linear profile $I_p$ and $T$ are constants. The translation stage must make a suitable number of trips along the fiber. $I_p$ should be close to the size of the heated region. Therefore, there has to be enough flame trips so that the hotzone size in Figure 6(a) can be accurately defined during the pull. Too few trips will result in unprocessed regions along the fiber. However, if there are too many trips, at the turning point of the trip, there will be overlap between each trip and the previous one and this can cause the overlapped area of fiber to be locally overtapered, meaning that an inhomogeneous taper would be achieved. In most cases, we have found that around 80 trips are sufficient, as also determined in other work.[32]

As an example, let us consider the tapers presented in Figure 6. The total pulling length of Taper 1 in Figure 6 was $d_0 = 55.6$ mm, calculated from Equation 9 and the number of trips selected was $N = 80$. This gives $I_p = 695$ $\mu$m. The hotzone size was then taken from the curve in Figure 6(a) for each 695 $\mu$m of elongation, i.e., $N \times I_p = d_0$. For each trip of the flame, a corresponding hotzone size, $L_i$, is now known, where $i = 0, 1, 2, 3, ...$ is the trip number and $i = 0$ represents the initial starting position and velocity of the flame. $P_r$ was set to 0.1 mm/s (0.05 mm/s per stage) which gives $T = 6.9$ s. The pull rate was selected to ensure that the maximum speed of the flame did not exceed 4 mm/s. For the speed of the flame, the hot zone size at each multiple of $I_p$ was divided by $T$. This yields a table of flame speed against hotzone size for each flame trip.

Next, a position coordinate, $P$, for the turning point of the flame translation stage must be made for each trip of the flame. It is clear that the distance the flame travels during each trip is equal to the hotzone size for that trip. The center of the taper is defined as the zero position. The starting position of the flame is equal to half of the initial hotzone, e.g. if $L_{i=1} = 30$ the flame starts at a position $P_{i=0} = -15$ and the initial velocity is zero, then $P_{i+1} = [P_i + ((-1^i) \times L_{i+1})] - \alpha$, where $\alpha = L_{i=1}/2$. The resulting flame position and speed for each flame trip is plotted in Figure 7 for Taper 1 and Taper 2.

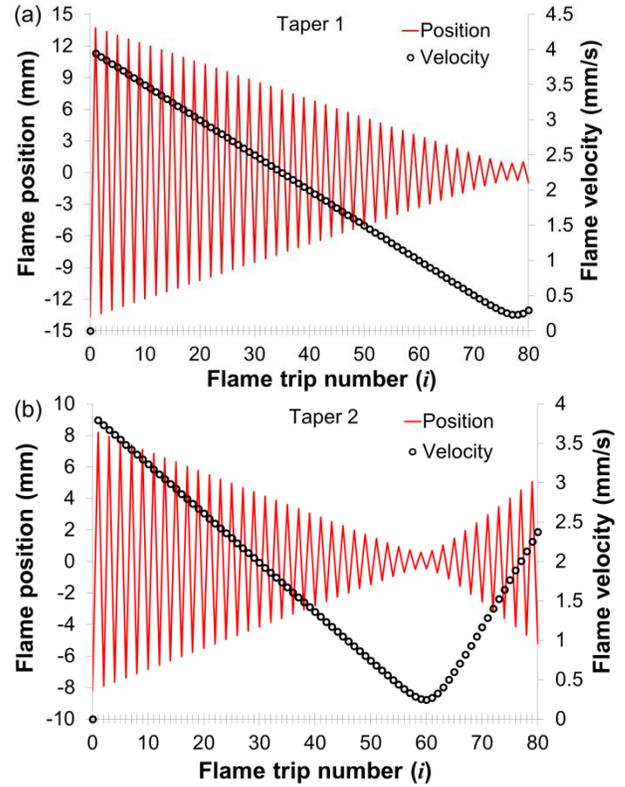

FIG. 7. The speed and position coordinates for the flame brushing stage plotted for each flame trip for (a) Taper 1 where $P_r = 0.1$ mm/s, $T = 6.9$ s, $I_p = 695$ $\mu$m, $d_0 = 55.6$ mm and (b) Taper 2 where $P_r = 0.125$ mm/s, $T = 4.32$ s, $I_p = 540$ $\mu$m, $d_0 = 43.2$ mm

MNFs with the profiles shown in Figure 6 were produced by our fiber pulling rig using the values in Figure 7. Their sizes were measured under an optical microscope. Figure 8 shows a comparison between the experimentally obtained profiles and the expected ones. Deviant data points are most likely due to errors in measurement with the microscope. There also appears to be a slight unexplained asymmetry in the profile of Taper 2; however, the overall fit is quite close. The transmitted power for both tapers was around 98%.

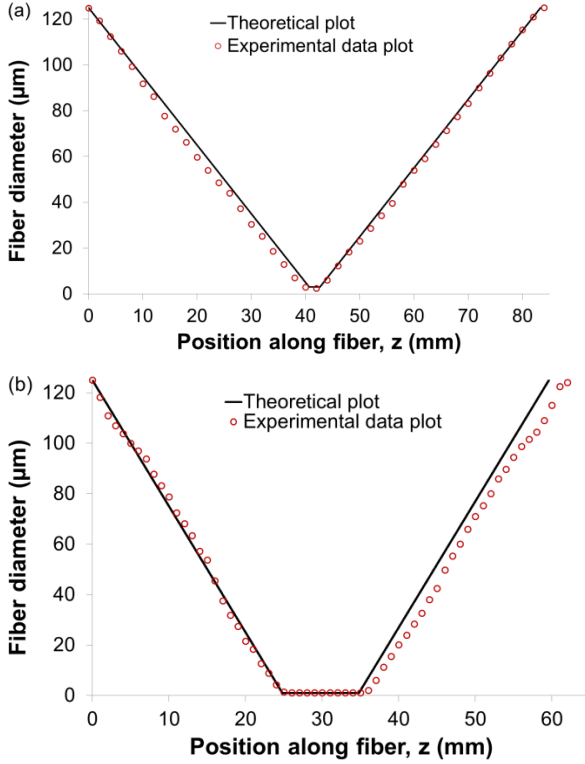

FIG. 8. The measured taper profiles for and the expected profiles for (a) Taper 1 and (b) Taper 2

## C. Tapered Fibers for Higher Mode Propagation

The propagation of higher order modes (HOMs) in optical waveguides is an interesting research topic for many applications, with our focus lying in the area of cold atom research. It has been shown that more cold atoms interact with the evanescent field around an MNF if HOMs propagate through the fiber as opposed to the fundamental mode alone.[60] This arises due to the increased spatial extension of the HOM field into the atom cloud. Kumar et al.[60] demonstrated that more atoms absorb light from the evanescent field of the higher order mode and, additionally, that the spontaneous fluorescence from resonantly excited atoms seems more likely to pump the higher order guided modes in the MNF, as predicted by Masalov and Minogin.[58] Interference between HOMs can be used to generate trapping sites for atoms,[59] thereby facilitating the development of neutral atom-based quantum technologies. With this in mind, the fabrication of an adiabatic taper profile for the higher order, $LP_{11}$, fiber mode at a wavelength of 780 nm was studied. The choice of wavelength was based on the cooling transition for $^{87}$Rb.

From Maxwell's equations for a cylindrical waveguide,[17] we can define the V-parameter such that:

$$V = \frac{2\pi r}{\lambda}\sqrt{n_1^2 - n_2^2}. \qquad (10)$$

Here, $r$ is the fiber radius, $\lambda$ is the wavelength of the input light and $n_1$ and $n_2$ represent the refractive indices of the core and the cladding of the fiber, respectively. In general, a fiber can support only one mode if the V-number is less than 2.405 and, in this case, the fiber is said to be single-mode. However, a single-mode fiber for one specific wavelength can transmit more than one mode if the propagating wavelength is shorter than the cutoff wavelength. For example, the Thorlabs SM1250-80 fiber, which was used in all the following work, has a cladding diameter of 80 μm. It is single-mode for 1300 nm, but it supports four families of modes ($LP_{01}, LP_{11}, LP_{21}, LP_{02}$) at the shorter wavelength of 780 nm (see Figure 9).

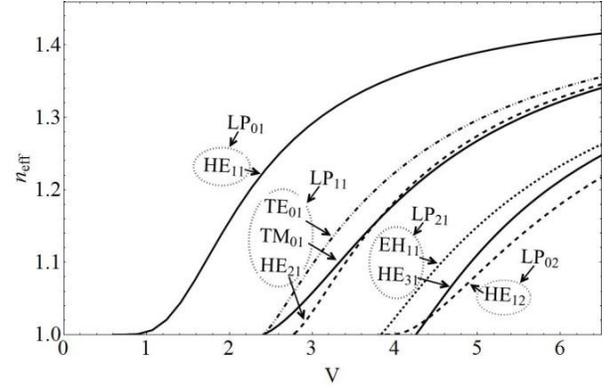

Fig.9. V-number plot for the first four modes in an 80 μm few mode fiber at a wavelength of 780 nm. Here, $n_{eff}$ is the effective refractive index of the propagating mode.

Due to the research interests mentioned above and theoretical/experimental knowledge from previous work,[33] the following discussion only considers propagation of the $LP_{11}$ higher order mode through the MNF.

There are many ways of generating an $LP_{11}$ mode in an optical fiber, such as generating a first order Bessel beam using a birefringent crystal and injecting it into a few mode fiber,[57] or using a phase plate to create a two lobed beam that corresponds to the $TEM_{01}$ mode in free space,[36] which, when launched into the fiber, excites the $LP_{11}$ mode. Alternatively, a Laguerre-Gaussian beam (LG) can be created using a spatial light modulator (SLM) and the LG mode can be coupled from free space into a few mode fiber.[33] The SLM works dynamically and can be used to efficiently obtain any desired beam profile with high beam quality. The liquid crystal on silicon spatial light modulator (SLM, Holoeye Pluto) used here is a reflective, phase only SLM. It has a surface resolution of 1920x1080 pixels and a fill factor of 87% with an image refresh rate of 60 Hz. A combination of computer generated phase discontinuity of the LG phase with a blaze grating was applied to the SLM. A linearly polarized light beam at a wavelength of 780 nm from a tunable Ti-Sapphire laser (Coherent, MBR-110) was incident onto the surface of the SLM.

The reflected output beam from the SLM was a high quality donut-shaped $LG_{11}$ beam from the first diffraction order. By choosing the proper objective lens, 50% coupling efficiency to the $LP_{11}$ mode of the few mode fiber was achieved. Both before and after coupling to the fiber, a CCD camera was used to image the beam profile. The resulting beam profile at the input to and output from the fiber is shown in Figure 10. The output consists of two lobes, as is characteristic of the $LP_{11}$ mode and it contains the $TE_{01}, TM_{01}$ and the coupled $HE_{21}$ modes (see Figures 9 and 10). Note that except for the $LP_{11}$ mode, there is some small amount of unexpected residual fundamental mode. This is mainly due to the quality of the generated beam and the efficiency of fiber coupling. By tapering the fiber until it breaks, the cutoff of all higher modes and the remaining portion of the residual fundamental mode can be observed. Subtracting the percentage of the contained fundamental mode power from the total power, a beam purity of 95% for the higher order mode ($LP_{11}$) in this few mode fiber was obtained. Another useful method for monitoring the mode beating is to take a Fourier transform of the transmitted signal. The spectrogram of the Fourier analysis reveals the beating between the different modes.[28,35]

First, the transmission of the $LP_{11}$ mode during the fabrication of a tapered fiber with an exponential profile was studied. Unlike for the fundamental fiber mode, the taper angle dependence is much stronger for the $LP_{11}$ mode, i.e., low loss occurs for shallower angles. For an exponential taper, a larger hotzone corresponds to a shallower taper angle, which ensures an adiabatic taper profile. However, this also means a longer taper. Therefore, choosing too large a hotzone can cause the fiber to sag under its own weight or to be easily broken. A compromised hotzone of 7 mm, pulling speed 0.125 mm/s, and flame moving speed of 0.4 mm/sec were applied. Pulling continued until the fiber broke and the transmitted power is plotted in Figure 10.

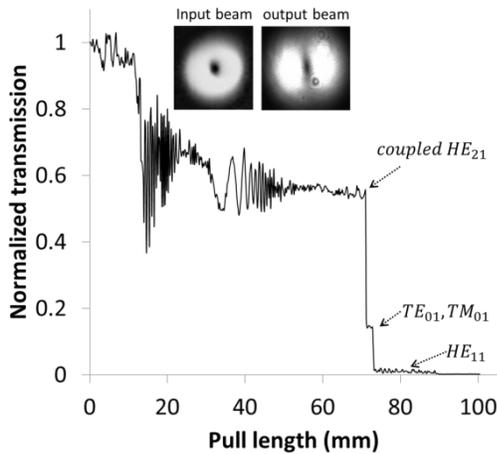

FIG.10. Transmission of the higher order mode in a taper with an exponential profile. Inset: a picture of the input and output beam profiles. The output exhibits the characteristic double lobes of a higher order mode.

For the exponential taper in Figure 10, a maximum transmission of 55% was obtained for the $LP_{11}$ mode. There is a sudden drop in the transmitted power at the start of the pull which is due to the fact that the local taper angle is not adiabatic for the higher order mode. To overcome this nonadiabaticity, initially a very shallow taper angle needs to be applied to avoid mode beating, but the overall taper length also needs to be considered. For this reason, a fiber was tailored so as to have a double linear taper profile. In this way, at the region where loss was strongest, a shallow taper angle of 0.5 mrad ($\Omega_1$) was applied. At the region where the mode coupling was not dominant, the fiber was tapered with a slightly larger angle of 1 mrad ($\Omega_2$). The length and diameter of the waist of the nanofiber was set to be 2 mm and 700 nm respectively. This preset diameter is just beyond the cutoff of the $LP_{11}$ mode (at around 660 nm) in the few mode fiber used.

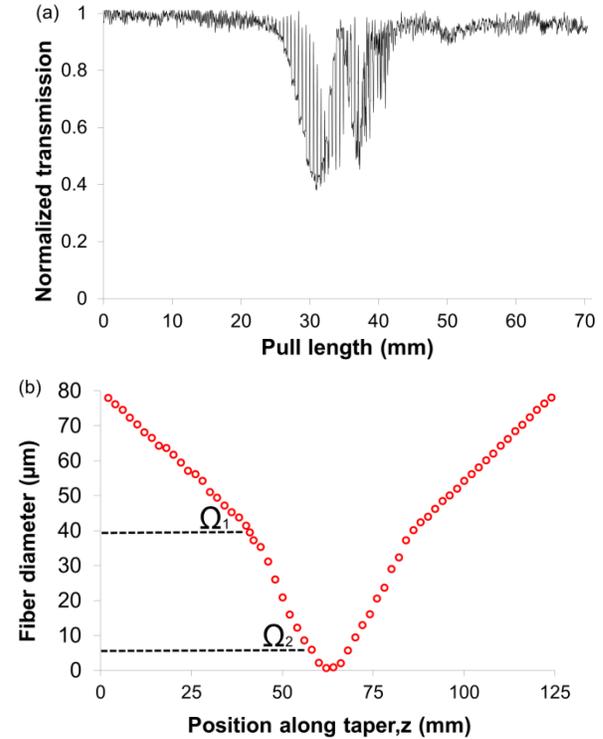

FIG.11. $LP_{11}$ mode transmission plot for double linear taper (left) and measured taper shape (right)

Figure 11 shows the transmission through the linear taper fiber. In the middle of the pulling there were some power fluctuations due to strong mode beating between core and cladding modes. Since the tapering angle is still small the cladding modes are eventually eliminated. $LP_{11}$ is confined at the cutoff waist diameter with an overall transmission of ~95% for the taper shown (Figure 11). Typically transmissions between 85% and 95% were achieved. These are the highest reported transmissions for the $LP_{11}$ modes in a fiber with an initial diameter of 80 µm. Fatemi *et al.* achieved 97% transmission for the $LP_{11}$ modes in a 50 µm fiber.

The profile of the nanofiber was determined using a Leica (Leica DM4000 M LED) optical microscope with a x100 objective. This system has a software package that allows digital zoom of the image and an estimated measurement of submicron structures. However, this is not sufficient to measure any variation in the diameter of the waist. Therefore, we used a scanning electron microscope

(SEM) for this measurement. We selected one fiber and measured its diameter along the length of its waist and observed no significant variation, confirming that the profile of the waist of the nanofiber was within the preset profile range. The linearity of the bi-taper fiber profile measured by the optical microscope showed that the measured taper angles $\Omega_1$ and $\Omega_2$ are 0.558 mrad and 1.051 mrad, respectively.

## V. CONCLUSION

In conclusion, we have reviewed and discussed the procedures and components necessary to build a flame brushing optical MNF pulling rig system, with exponential or linear taper profiles. Test measurements were performed using the rig system and very high transmissions of the fundamental and the $LP_{11}$ higher order mode have been demonstrated. This system is an excellent tool for producing complex nanofiber shapes with high accuracy and uniformity. One can switch the taper profile between linear and exponential, and selectively change the linear taper angle to any desired angle. On both exponential and linear tapers, one can control the fiber size with good accuracy and reproducibility. A laser wavelength of 780 nm was used in the test measurements - creating adiabatic tapers for different wavelengths is also possible. We are currently integrating the few mode optical nanofibers into a cold sample of $^{87}$Rb in order to study atom interactions with higher order modes.

## ACKNOWLEDGEMENTS

The authors wish to thank Alan Curtis for early contributions to building the fiber rig and Mary Frawley for useful discussions on higher order mode propagation. This work is funded through OIST Graduate University.